\documentclass[12pt]{iopart}
\usepackage{graphicx}

\begin{document}
\def \ee {\varepsilon}
\thispagestyle{empty}
\title[]{
Van der Waals and Casimir interactions between atoms and
carbon nanotubes
}

\author{G~L~Klimchitskaya,${}^{1,2}$ E~V~Blagov${}^3$
and V~M~Mostepanenko${}^{1,3}$}

\address{${}^1$Center of Theoretical Studies and Institute for Theoretical 
Physics, Leipzig University,
D-04009, Leipzig, Germany}

\address{$^2$North-West Technical University, Millionnaya St. 5,
St.Petersburg, 191065, Russia}

\address{${}^3$
Noncommercial Partnership  ``Scientific Instruments'', 
Tverskaya St. 11, Moscow, 103905, Russia}

\begin{abstract}
The van der Waals and Casimir interactions of a hydrogen atom
(molecule) with a single-walled and a multiwalled carbon
nanotubes are compared. It is shown that the macroscopic
concept of graphite dielectric permittivity is already
applicable for nanotubes with only two or three walls.
The absorption of hydrogen atoms by a nanotube at separations
below one nanometer is considered. The lateral force due to
exchange repulsion moves the atom to a position above the cell
center, where it is absorbed by the nanotube because the
repulsive force cannot balance the van der Waals attraction.
\end{abstract}

\section{Introduction}

Considerable recent attention has been focused on applications of
the van der Waals and Casimir forces in nanotechnology. When the
characteristic sizes of microdevices shrink below a micrometer, the
collective quantum phenomena caused by the existence of zero-point
oscillations of the electromagnetic field come into play. 
At separations below 100\,nm the role of physical phenomena originating
from vacuum oscillations (primarily of the van der Waals and Casimir
forces) can match the role of characteristic electric force and may
even become dominant. This was first stressed long ago in \cite{1}
and was commonly accepted in the beginning of the Third Millenium. 
By then several experiments on measurement of the Casimir force
between metallic surfaces had already been performed \cite{1a}. 
This laid the groundwork for the application of dispersion forces
(which is a generic name for the van der Waals and Casimir force) in
nanomechanical devices \cite{2,3,4}, noncontact atomic friction
\cite{5,6,7,8,8a}, carbon nanostructures \cite{9,10,11} and related 
subjects. Thereafter, the precision of measurements of the Casimir
force was significantly increased \cite{12} and different methods to
control the force magnitude were elaborated \cite{13} opening
possible applications to nanotweezers, nanoscale actuators
and other nanomachines.

Dispersion interaction is also of much importance for the understanding
of absorption phenomena of atoms and molecules by nanostructures, 
This subject is currently topical in connection with the problem of 
hydrogen storage in carbon nanotubes. Until recently it was studied
using the phenomenological density-functional theory (see, e.g.,
\cite{10a,9,11}). In \cite{14,15} the Lifshitz theory of the van der Waals
and Casimir force \cite{17a}
was applied for the cases of an atom (molecule)
interacting with a plane surface of a graphite plate or multiwall
carbon nanotube. This was achieved by using 
a classical model of the frequency dependent
dielectric permittivities of a uniaxial crystal and proximity force
approximation \cite{16}.

Some properties of
the monoatomic sheet of C atoms (graphene) admit a simplified
model description in
terms of the two-dimensional free electron gas. In doing so the graphene
sheet is characterized by some typical wave number $K$ determined
by the parameters of the hexagonal structure of graphite. In \cite{17}
the interaction of the electromagnetic oscillations with such a sheet
was considered and the reflection coefficients were found. In \cite{18}
the Lifshitz-type formula was obtained for the van der Waals and
Casimir interaction between the two parallel plasma sheets. Using this
model in \cite{19} the interaction between graphene and a material plate, 
graphene and an atom or a molecule, and between a single-walled carbon
nanotube and a material plate was also described by means of the
Lifshitz-type formulas. Finally, in \cite{20} the Lifshitz-type formula
was obtained for the van der Waals and Casimir interaction between an atom
(molecule) and a single-walled carbon nanotube. All above mentioned
results are applicable if the separation distances are sufficiently large 
(typically larger than 1\,nm).

In this paper, we compare the van der Waals interactions of a hydrogen
atom (molecule) with a single-walled and a multiwalled carbon nanotube.
The former is approximately 
modeled as a cylindrical plasma sheet, whereas the
latter as a cylindrical shell of finite thickness characterized by
graphite dielectric permittivities for the ordinary and extraordinary
rays. We arrive at the conclusion that the macroscopic concept of
dielectric permittivity is already applicable for nanotubes
containing only two or three walls (Section 2). We also consider
the interaction of atoms or molecules with carbon nanotubes at
separations below 1\,nm where the Lifshitz-type formulas obtained
in \cite{18,19,20} are not applicable. Using the method of
phenomenological potentials and disregarding the role of
chemical forces, we find that at separations below 1\,nm
the exchange repulsion gives rise to the lateral force that moves hydrogen 
atoms towards the cell center. In the position above a cell center,
the repulsive force cannot balance the van der Waals attraction.
As a result, the atom penetrates inside the nanotube (Section 3). 
This effect is analogous to the breaks of constant force
surfaces that arise when scanning the monoatomic tip of an atomic
force microscope above a closely packed lattice in contact mode \cite{21,22}.
The discussion of the obtained results and conclusions are contained
in Section 4.

\section{Comparison of atom-nanotube interaction in the cases of
multiwalled and single-walled nanotubes}

The free energy and force between an atom (molecule) and a multiwalled
or single-walled carbon nanotube separated by a distance $a$ can be
represented in the form
\begin{equation}
{\cal F}(a,T)=-\frac{C_3(a,T)}{a^3}, \qquad
F(a,T)=-\frac{C_F(a,T)}{a^4}.
\label{eq1}
\end{equation}
\noindent
Here, $T$ is the temperature of graphite which is supposed to be in
thermal equilibrium with the surroundings.
The coefficients $C_3(a,T)$ and $C_F(a,T)$ are defined in such a way
that at short separations equation (\ref{eq1}) reproduces the
nonrelativistic van der Waals interaction (in this limit $C_3$ and 
$C_F$ do not depend on separation and temperature and it holds $C_F=3C_3$).
The Lifshitz-type formula for the coefficient $C_3(a,T)$ was obtained
in \cite{14,20} by using the proximity force approximation \cite{16,23}
\begin{eqnarray}
&&
C_3(a,T)=\frac{k_BT}{8}\sqrt{\frac{R}{R+a}}\left\{
\frac{4R+3a}{2(R+a)}\alpha(0)\right.
\label{eq2} \\
&&\phantom{aaaaa}
+\sum_{l=1}^{\infty}\alpha(i\zeta_l\omega_c)
\int_{\zeta_l}^{\infty}dy\,y\,{\rm e}^{-y}\left[y-\frac{a}{2(R+a)}\right]
\nonumber \\
&&\phantom{aaaaa}
\times\left.\left[2r_{\rm TM}({\rm i}\zeta_l,y)+\frac{\zeta_l^2}{y^2}
\Bigl(r_{\rm TE}({\rm i}\zeta_l,y)-
r_{\rm TM}({\rm i}\zeta_l,y)\Bigr)\right]\right\}.
\nonumber
\end{eqnarray}
\noindent
Here, $R$ is the nanotube radius, $\alpha(\omega)$ is the dynamic 
polarizability of an atom or a molecule, $k_B$ is the Boltzmann constant
and the dimensionless Matsubara frequencies $\zeta_l$ are connected with
the dimensional ones by the equalities
\begin{equation}
\zeta_l=\frac{\xi_l}{\omega_c}, \qquad
\xi_l=2\pi\frac{k_BT}{\hbar}l, \qquad
\omega_c=\frac{c}{2a}, 
\qquad l=0,\,1,\,2,\,\ldots\, .
\label{eq3}
\end{equation}
In our model description,
the reflection coefficients for the two independent polarizations of the
electromagnetic field, $r_{\rm TM}$ and $r_{\rm TE}$, have different
forms in the cases of multiwalled and single-walled carbon nanotubes.
For multiwall nanotubes, they are expressed in terms of two dissimilar
dielectric permittivities of graphite, 
$\varepsilon_x(\omega)=\varepsilon_y(\omega)$ and $\varepsilon_z(\omega)$,
where the crystal optical axis $z$ is perpendicular to a cylindrical
surface
\begin{eqnarray}
&&
r_{\rm TM}^{\rm mw}({\rm i}\zeta_l,y)=
\frac{\varepsilon_{xl}\varepsilon_{zl}y^2-
f_z^2}{\varepsilon_{xl}\varepsilon_{zl}y^2+f_z^2+
2\sqrt{\varepsilon_{xl}\varepsilon_{zl}}yf_z
{\rm coth}[f_zd/(2a)]},
\nonumber \\
&&
r_{\rm TE}^{\rm mw}({\rm i}\zeta_l,y)=
\frac{f_x^2-y^2}{f_x^2+y^2+2yf_x{\rm coth}[f_xd/(2a)]}.
\label{eq4}
\end{eqnarray}
\noindent
Here, the thickness of a nanotube is $d=3.4(N-1)\,$\AA, where
$N$ is the number of walls, and the following notations are introduced:
\begin{eqnarray}
&&
\varepsilon_{xl}=\varepsilon_x({\rm i}\zeta_l\omega_c), \qquad
\varepsilon_{zl}=\varepsilon_z({\rm i}\zeta_l\omega_c),
\label{eq5} \\
&&
f_z^2=y^2+\zeta_l^2(\varepsilon_{zl}-1), \qquad
f_x^2=y^2+\zeta_l^2(\varepsilon_{xl}-1).
\nonumber
\end{eqnarray}
\noindent
The dielectric permittivities of graphite along the imaginary frequency
axis are calculated in \cite{14} on the basis of tabulated optical
data from \cite{24}.

For single-walled nanotubes, we use the simplified model description by
means of a cylindrical plasma sheet. Then
the reflection coefficients are expressed in
terms of the wave number of the sheet, $K$, which is determined by the
density $n$ of $\pi$ electrons. Bearing in mind that there are two
$\pi$ electrons per one hexagonal cell with a side length $r_0=1.421\,$\AA,
we arrive at
\begin{equation}
K=2\pi\frac{ne^2}{mc^2}=6.75\times 10^{5}\,\mbox{m}^{-1},
\qquad
n=\frac{4}{3\sqrt{3}r_0^2},
\label{eq6}
\end{equation}
\noindent
where $e$ and $m$ are the electron charge and mass, respectively.
The resulting coefficients are
\begin{equation}
r_{\rm TM}^{\rm sw}({\rm i}\zeta_l,y)=\frac{2yaK}{2yaK+\zeta_l^2}, \qquad
r_{\rm TE}^{\rm sw}({\rm i}\zeta_l,y)=\frac{2aK}{2aK+y}.
\label{eq7}
\end{equation}

The Lifshitz-type formula for the coefficient $C_F(a,T)$ 
in (\ref{eq1}) can be presented using the same notations \cite{14,20}. 
We emphasize that within the considered models
equation (\ref{eq1}) with the coefficients $C_3$ 
and $C_F$ is applicable at both 
short and large separations, i.e., both in nonrelativistic and
relativistic regimes, and also in the transition region between the two 
regimes. We, however, consider $a>1\,$nm because at shorter separations 
the atomic structure of nanotube wall and also other forces of different
physical nature in addition to dispersion interaction should be taken 
into account.  In the framework of used models, the error introduced 
from the application of
the proximity force approximation is less that 1\% within the separation
region from 0 to $R/2$ \cite{23}.
It is notable also that the above models do not take into account
nanotube chirality. This can be included by using the optical data
for nanotube complex index of refraction. We would like to underline
that the used models, especially the model of a cylindrical plasma
sheet in application to single-walled carbon nanotubes, do not claim
a complete description of all nanotubes properties. Specifically, it
remains unclear if it possible to describe nanotubes with surfaces
like metals and like semiconductors by varying only one parameter
$K$ in the reflection coefficients (\ref{eq7}).

To perform computations using above equations in the case of hydrogen
atoms and molecules, one neads the explicit expressions for the
atomic and molecular dynamic polarizabilities. As was shown in \cite{14,29},
the dynamic polarizability can be represented with sufficient precision
using the single-oscillator model
\begin{equation}
\alpha_{a}(i\xi_l)=\frac{g_a}{\omega_a^2+\xi_l^2}, \qquad
\alpha_{m}(i\xi_l)=\frac{g_m}{\omega_m^2+\xi_l^2}
\label{eq9}
\end{equation}
\noindent
for hydrogen atom and molecule, respectively. Here,
$\alpha_a(0)=4.50\,$a.u., $\omega_a=11.65\,$eV,  
$\alpha_m(0)=5.439\,$a.u., $\omega_a=14.09\,$eV.

We have performed computations of the van der Waals free energy of
hydrogen atoms and molecules and multiwalled and single-walled carbon
nanotubes modeled as discussed above. 
For multiwalled nanotubes, equations (\ref{eq1}), (\ref{eq2}),
(\ref{eq4}), (\ref{eq9}) were used. For single-walled
nanotubes equation (\ref{eq4}) was replaced with (\ref{eq7}).
The computational results for the coefficient $C_3$ as a function of
the number of walls are presented as solid dots in figure 1(a) (hydrogen
atom) and in figure 1(b) (hydrogen molecule) for nanotubes with
$R=5\,$nm. In both cases the solid dots connected with a solid line are
related to a multiwalled nanotube and the solid dots marked by 1, 2, and 3
are related to a single-walled nanotube for separation distances of
$a=1,\,2$, and 3\,nm, respectively.   
\begin{figure*}[t]
\vspace*{-11.5cm}
\hspace*{-0.5cm}\includegraphics{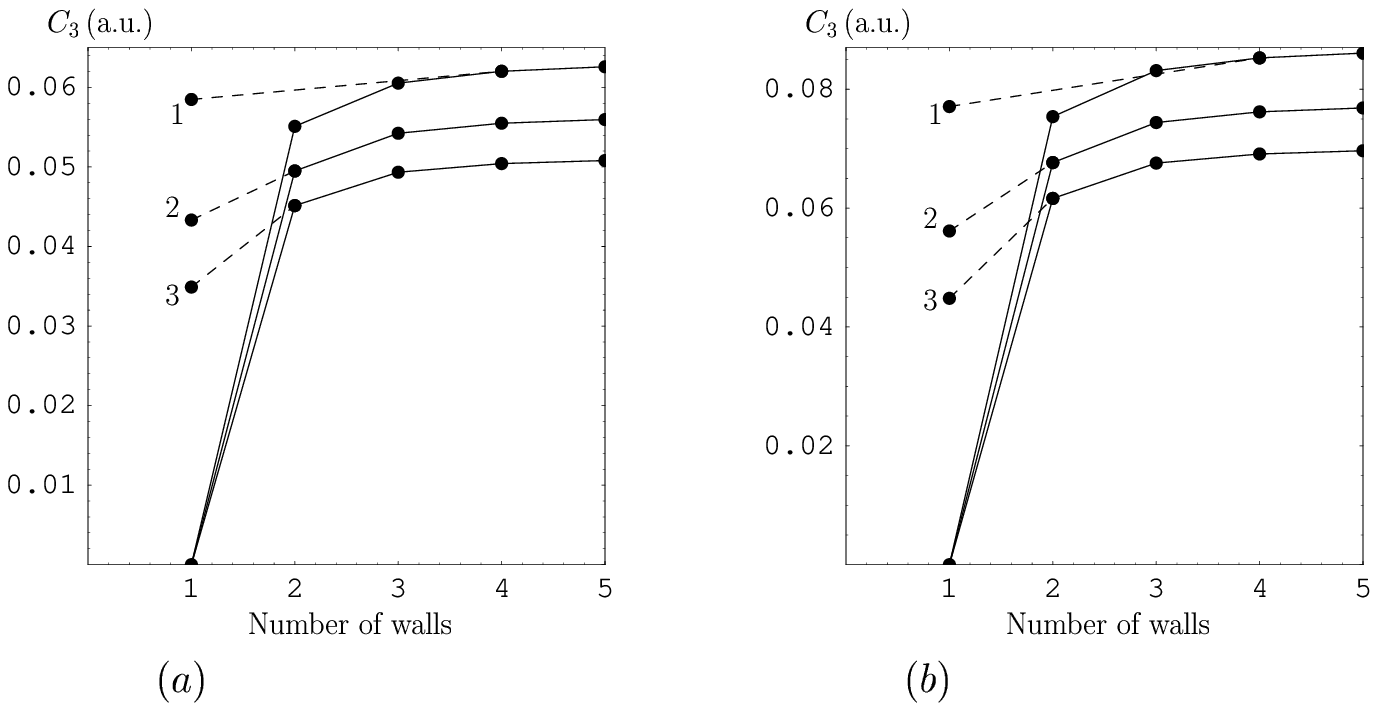}
\vspace*{-12.cm}
\caption{
The van der Waals coefficient as a function of the number of walls
for hydrogen atom (a) and molecule (b) interacting with the
multiwalled carbon nanotube (solid dots connected with solid
lines) and with the single-walled carbon nanotube (solid dots
1, 2 and 3) both of $R=5\,$nm radius spaced at 1, 2, and 3\,nm
from the atom (a) or molecule (b). The thicknesses of multiwalled 
nanotubes with 2, 3, 4 and 5 walls are 0.34, 0.68, 1.02 and 1.36\,nm,
respectively.
}
\end{figure*}
For a ``multiwalled'' nanotube with only one wall $C_3=0$, as expected, 
because in this case nanotube thickness $d=0$ and hence the reflection
coefficients (\ref{eq4}) vanish. As is seen in figure 1(a),(b), the
magnitudes of the van der Waals coefficients $C_3$ for a multiwalled
nanotube with two walls at a separations of 2 or 3\,nm from an atom
or a molecule are in correct relation to the magnitudes computed for
a single-walled nanotube (solid dots marked 2 and 3). Thus, the
macroscopic concept of graphite dielectric permittivity is already
applicable for nanotubes with only two or three walls provided the
separation distance to an atom or a molecule is larger than 2\,nm.
For nanotubes with three walls the same holds for sepataions
of 1\,nm between an atom or a molecule and a nanotube.
These results are expected because it is known that surface corrections
to local fields in ordinary crystals generally become negligible
already in the third or fourth lattice plane.
Similar results but with other numerical values are also valid for
the force coefficient $C_F$. To give an example, for nanotubes
of $R=5\,$nm radius at $a=1\,$nm from a hydrogen atom 
$C_3=0.0585\,$a.u. but $C_F=0.197\,$a.u. At the same separation
but for nanotubes of $R=2\,$nm radius,
$C_3=0.0503\,$a.u. and $C_F=0.175\,$a.u.

\section{Absorption of hydrogen atoms by carbon nanotubes}

The above formalism based on the Lifshitz-type formulas permits to
calculate the attractive van der Waals and Casimir force acting between
both multiwalled and single-walled carbon nanotubes and hydrogen atoms or
molecules down to 1\,nm separation distance using the simplified models 
described above. 
However, at shorter separations
the atomic structure of nanotube and some other physical interactions
in addition to dispersion forces should be taken into account.
Because of this, 
complete calculation of absorption power of carbon nanotubes
requires a detailed investigation of the microscopic interaction
mechanisms including chemical forces and short-range exchange forces.
The comparative role of different forces in absorption process has not
been yet investigated. Below we consider the role of dispersion and
exchange forces at separations less than 1\,nm.
We argue that the combined action of the interatomic attractive
van der Waals force and of the repulsive exchange force,
taken into account using a simple model, leads to the
absorption of hydrogen atom by a nanotube. The investigation of the role
of chemical forces is an interesting problem to be solved in future.

The exchange repulsion between C and H atoms at a separation $r$ apart
can be approximately described by means of the phenomenological potential
$U_{\rm rep}(r)=\alpha/r^{12}$ \cite{30}, where $\alpha$ is some
coefficient. Papers \cite{21,22} investigated the scanning of the
monoatomic tip of an atomic force microscope along the constant force 
surface above the closely packed crystal lattice in contact mode.
It was supposed that the repulsive potential $U_{\rm rep}(r)$
is the single factor to be taken into account. As was shown in \cite{21,22},
if the initial height of the tip $h$ satisfies the condition
$h/r_0>0.61$ (here, $r_0$ is the equilibrium interatomic distance in the 
lattice), the constant force surface under consideration is continuous.
If, however, the initial height is sufficiently small ($h/r_0\leq 0.61$),
the constant force surfaces have breaks above the regions between crystal
atoms. For graphite lattice the atom in the center of a hexagonal 
cell is missing  which only increases the areas of the breaks.

Let us consider now the hexagonal lattice shown in figure 2(a) with a
hydrogen atom at a height $h$ above it in the positions between points A 
and B (the lattice parameter is $r_0$). There is the lateral force acting 
on an atom at a height $h$ directed to a cell center. This can be seen in
the following way. Let us consider the total repulsive energy due to
interaction with a potential $U_{\rm rep}$ with all nearest
C atoms with coordinates $(x_i,y_i,0)$, e.g., with 4 atoms in the
position A and 6 atoms in the position B (the first coordinative sphere).
This energy is given by
\begin{equation}
E_{\rm rep}(x,h)=\sum_{i}\frac{\alpha}{\left[(x-x_i)^2+y_i^2+h^2\right]^6},
\label{eq10}
\end{equation}
\noindent
where $x=0$ in the position A and $x=r_0$ in the position B.
In figure 2(b) we plot the ratio 
$E_{\rm rep}(x,h)/E_{\rm rep}(0,h)$ as a function of $(r_0-x)/r_0$ for
different heights:
$h=1.5r_0$ (line 1), $h=r_0$ (line 2) and $h=0.5r_0$ (line 3).
As is seen in figure 2(b), the equilibrium position of an atom at $x=0$
(point A) is unstable, whereas the equilibrium position of an atom
at $x=r_0$ (point B) is stable. Thus, atom will be displaced by a
horizontal force to the position B above the cell center (in fact
it should be displaced to the center of one of the three hexagonal
cells around a position A).
\begin{figure*}[t]
\vspace*{-14.4cm}
\hspace*{-0.5cm}\includegraphics{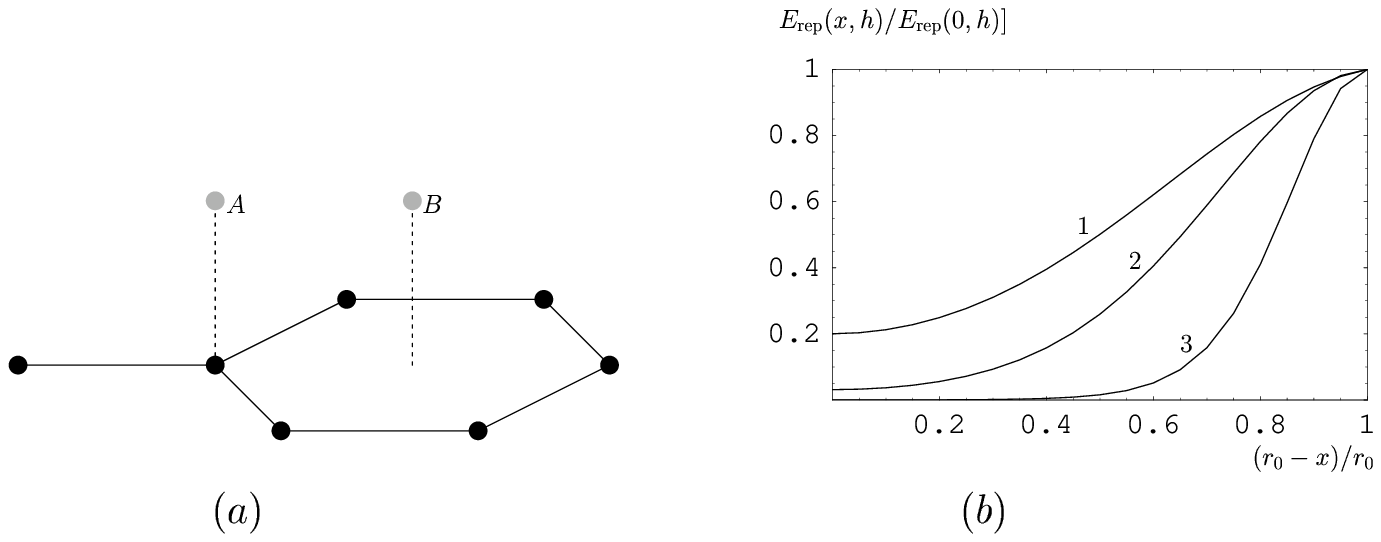}
\vspace*{-11.cm}
\caption{
(a) Hydrogen atoms in the positions A and B at a height $h$ above
the hexagonal lattice of C atoms. (b) The relative energy of exchange 
repulsive interaction between the hydrogen atom at a height $h$ above
the hexagonal lattice of C atoms and neighboring C atoms as a
function of $(r_0-x)/r_0$. Here $x$ is the lateral displacement from
the C atom below a point A to the center of the cell.
}
\end{figure*}

We further suppose that the total interaction of H atom with C atom at
a separation $r$ below 1\,nm is described by the Lennard-Jones
interaction potential
\begin{equation}
U_{\rm LJ}(r)=\frac{\alpha}{r^{12}}-\frac{\beta}{r^6},
\label{eq11}
\end{equation}
\noindent
where $\beta$ is the constant of interatomic van der Waals interaction.
The equilibrium separation distance $h_0$ between one C and one H atom
satisfies the condition
\begin{equation}
-\left.\frac{\partial U_{\rm LJ}(r)}{\partial r}\right|_{r=h_0}=
\frac{12\alpha}{h_0^{13}}-\frac{6\beta}{h_0^7}=0.
\label{eq12}
\end{equation}
\noindent
This leads to the relation $\beta=2\alpha/h_0^6$.

The total force acting on an H atom at a height $h$ in the position B
is determined by the six neighboring C atoms. It is equal to the sum of
vertical projections of forces with potential (\ref{eq11})
\begin{equation}
F_B=\frac{72\alpha h}{(r_0^2+h^2)^7}-
\frac{36\beta h}{(r_0^2+h^2)^4}.
\label{eq13}
\end{equation}
\noindent
The hydrogen atom would be absorbed by a nanotube if $F_B<0$ for any
value of $h$. Taking into account the above connection between
$\alpha$ and $\beta$, we obtain from (\ref{eq13})
\begin{equation}
\frac{1}{(r_0^2+h^2)^3}-
\frac{1}{h_0^6}<0.
\label{eq14}
\end{equation}
\noindent
This inequality is for sure satisfied if $h_0<r_0$.

It is common knowledge \cite{30,31} that the equilibrium position of
the two atoms
\begin{equation}
h_0\approx R_{ion}^{(1)}+R_{ion}^{(2)}\leq R_a^{(1)}+R_a^{(2)},
\label{eq15}
\end{equation}
\noindent
where $R_{ion}^{(k)}$ and $R_a^{(k)}$ are the so-called {\it ionic}
and {\it atomic} radii, respectively. For C and H atoms under consideration
it holds $R_a^{(C)}=0.77\,${\AA} and $R_a^{(H)}=0.53\,$\AA \cite{32}.
Thus, in our case
\begin{equation}
h_0\leq R_a^{(C)}+R_a^{(H)}=1.3\,\mbox{\AA}.
\label{eq16}
\end{equation}
\noindent
If to take into account that for graphite cell $r_0=1.42\,$\AA, we arrive
at the conclusion that the condition $h_0<r_0$ is satishied and hydrogen
atom will be absorbed by a nanotube.

\section{Conclusions and discussion}

In the above we have presented the Lifshitz-type formulas for the
van der Waals and Casimir free energy and force for the configuration of
hydrogen atoms or molecules in close proximity to multiwalled or
single-walled carbon nanotubes. Multiwalled nanotubes are 
approximately described
by graphite dielectric permittivity, whereas single-walled nanotubes are 
considered in the approximation of a two-dimensional gas of free electrons.
Both descriptions are of model character and do not claim complete
description of all nanotube properties. They are
applicable at separations larger than 1\,nm.
By comparing of calculation results for multiwalled nanotubes with
those for single-walled nanotubes, the conclusion was made that the
macroscopic description using the concept of dielectric permittivity
is already applicable for nanotubes with only two or three walls
depending on the separation distance to the hydrogen atom.

We have also considered separations below 1\,nm where the exchange repulsion 
plays an important role. 
By disregarding the role of chemical forces,
it was shown that under the influence of a lateral 
force originating from exchange repulsion, the hydrogen atom is moved towards 
a cell center. Simple analysis shows that at atomic distances above the
cell center the exchange repulsion cannot balance the van der Waals
attraction. As a result the hydrogen atom is absorbed by the nanotube.
This is analogous to the effect of breaks on the constant force
surfaces which arise when scanning of the monoatomic tip of an atomic
force microscope above a surface in the strong repulsive mode. In future
it would be interesting to consider the influence of the previously
absorbed atoms on the absorption process and to determine the
resulting absorption rate.

\section*{Acknowledgments}
VMM and GLK are
grateful to the Center of Theoretical Studies and Institute
for Theoretical Physics, Leipzig University for their kind
hospitality. They were supported by Deutsche Forschungsgemeinschaft,
Grant No.~436\,RUS\,113/789/0--3. 

\end{document}